\documentclass{amsart}

\usepackage{amsmath}
\usepackage{amsthm}
\usepackage{amsfonts}
\usepackage{dsfont}

\newcommand{\reals}{\mathbb{R}}

\newcommand{\bracketa}[1]{\big[#1\big]}
\newcommand{\bracketb}[1]{\Big[#1\Big]}
\newcommand{\bracketc}[1]{\bigg[#1\bigg]}

\newcommand{\pb}[1]{\left\{#1\right\}}
\newcommand{\com}[1]{\left[#1\right]}
\newcommand{\coma}[1]{\big[#1\big]}

\newcommand{\norm}[1]{\left|\left|#1\right|\right|}
\newcommand{\abs}[1]{\left|#1\right|}

\newcommand{\paraa}[1]{\big(#1\big)}
\newcommand{\parab}[1]{\Big(#1\Big)}
\newcommand{\parac}[1]{\bigg(#1\bigg)}

\newtheorem{theorem}{Theorem}[section]

\newtheorem{proposition}[theorem]{Proposition}

\theoremstyle{definition}
\newtheorem{definition}[theorem]{Definition}
\theoremstyle{remark}
\newtheorem{remark}[theorem]{Remark}
\numberwithin{equation}{section}

\newcommand{\trh}{\widehat{\operatorname{tr}}\,}
\newcommand{\Tr}{\operatorname{Tr}}

\renewcommand{\P}{\mathcal{P}}
\newcommand{\Ph}{\hat{\P}}
\renewcommand{\S}{\mathcal{S}}
\newcommand{\Sh}{\hat{\S}}
\newcommand{\ShA}{\hat{\S}_A}
\newcommand{\SA}{\S_A}

\newcommand{\xv}{\vec{x}}
\newcommand{\mv}{\vec{m}}
\newcommand{\nv}{\vec{n}}
\newcommand{\gb}{\,\bar{\!g}}

\renewcommand{\d}{\partial}
\newcommand{\TSigma}{T\Sigma}
\newcommand{\eps}{\varepsilon}

\newcommand{\nablab}{\bar{\nabla}}

\newcommand{\Rb}{\bar{R}}
\newcommand{\TN}{T^{(N)}}

\newcommand{\Ta}{T^{\alpha}}
\newcommand{\Na}{N_\alpha}
\newcommand{\Kh}{\hat{K}}
\newcommand{\chih}{\hat{\chi}}
\newcommand{\vphi}{\varphi}
\newcommand{\vphiv}{\vec{\vphi}}
\renewcommand{\mid}{\mathds{1}}
\newcommand{\gd}{g^\dagger}
\newcommand{\hd}{h^\dagger}
\newcommand{\limNinf}{\lim_{N\to\infty}}
\newcommand{\limainfty}{\lim_{\alpha\to\infty}}
\newcommand{\hbara}{\hbar_\alpha}

\renewcommand{\d}{\partial}
\newcommand{\Gammab}{\bar{\Gamma}}
\newcommand{\Gammah}{\hat{\Gamma}}
\newcommand{\gammah}{\hat{\gamma}}

\newcommand{\fh}{\hat{f}}
\newcommand{\Gh}{\hat{G}}
\newcommand{\ff}{f\!f}
\newcommand{\Wd}{W^\dagger}
\newcommand{\Ch}{\hat{C}}

\newcommand{\fha}{\fh_\alpha}
\newcommand{\fhua}{\fh^\alpha}

\title[Discrete curvature and the Gauss--Bonnet theorem]{Discrete curvature and the \\Gauss--Bonnet theorem}

\author{Joakim Arnlind}
\address{Max Planck Institute for Gravitational Physics (AEI), Am M\"uhlenberg 1, D-14476 Golm, Germany.}
\email{joakim.arnlind@aei.mpg.de}

\author{Jens Hoppe}
\address{Kungliga Tekniska
  H\"ogskolan, 10044 Stockholm, Sweden.}
\email{hoppe@math.kth.se}

\author{Gerhard Huisken}
\address{Max Planck Institute for Gravitational Physics (AEI), Am M\"uhlenberg 1, D-14476 Golm, Germany.}
\email{gerhard.huisken@aei.mpg.de}

\thanks{}

\subjclass[2000]{}

\begin{document}

\begin{abstract}
  For matrix analogues of embedded surfaces we define discrete
  curvatures and Euler characteristics, and a non-commutative
  Gauss--Bonnet theorem is shown to follow. We derive simple
  expressions for the discrete Gauss curvature in terms of matrices
  representing the embedding coordinates, and provide a large class of
  explicit examples illustrating the new notions.
\end{abstract}

\maketitle

\section{Introduction}

\noindent A particular way of discretizing surfaces by replacing
functions by matrices has for a long time been used in physics to
obtain a quantum theory of surfaces (membranes) moving in Minkowski
space, sweeping out 3-manifolds of vanishing mean curvature
\cite{h:phdthesis}. The discretization, sometimes called ``Matrix
Regularization'', is of independent mathematical interest and contains
many interesting structures. One of the main features of the
correspondence between functions on the surfaces and matrices is that
the Poisson bracket of two functions becomes the commutator of two
matrices. This allows for an easy construction of discrete analogues
of any expression involving products and Poisson brackets of
functions. In a recent paper \cite{ahh:psurface}, the geometry of
surfaces embedded in Riemannian manifolds has been expressed in terms
of Poisson brackets of the embedding coordinates. Using these
formulas, one can define discretizations of the Gaussian curvature and
the Euler characteristic, and it is immediate to prove a discrete
Gauss-Bonnet theorem (see Theorem \ref{thm:discreteEuler}).

Apart from being interesting in their own right, these discrete
concepts might also help to solve questions related to the
regularization in the above mentioned Membrane Theory. For instance,
solving the equations of motion in Membrane Theory yields matrices
corresponding to a discrete surface. As solutions corresponding to
surfaces of arbitrary topology exist, one would like to be able to
determine the geometry from the matrices in some way. In Theorem
\ref{thm:discreteCurvature} we provide formulas for computing the
discrete curvature and the discrete Euler characteristic given the
matrix analogues of the embedding coordinates (which are the solutions
to the equations of motion in Membrane Theory). Thus, in the limit of
large matrices one may determine the Euler characteristic, and hence
the topology, of the surface.

\section{Surface geometry and Poisson brackets}

\noindent Let us recall some of the results obtained in
\cite{ahh:psurface}. Consider a surface $\Sigma$ embedded in a
Riemannian manifold $M$, of dimension $m=2+p$, via the coordinates
$x^1(u^1,u^2),\ldots,x^m(u^1,u^2)$, where $u^1,u^2$ are local
coordinates on $\Sigma$. Furthermore, let
$n^1_A(u^1,u^2),\ldots,n^m_A(u^1,u^2)$ for $A=1,\ldots,p$ denote the
components of $p$ orthonormal vectors $N_A$ normal to the surface at each
point. Indices $i,j,k,l$ will run from $1$ to $m$ and indices $a,b$
will run from $1$ to $2$. The metric of $M$ is denoted by $\gb_{ij}$,
the Christoffel symbols by $\Gammab^i_{jk}$ and the covariant
derivative by $\nablab$. Regarded as a subspace of $TM$, the tangent
space $T\Sigma$ is spanned by the vectors $e_a=(\d_ax^i)\d_i$.

Letting $\rho(u^1,u^2)$ be an arbitrary non-vanishing density on
$\Sigma$, one defines a Poisson bracket on $C^\infty(\Sigma)$ by
setting
\begin{align}
  \{f,h\} = \frac{1}{\rho}\eps^{ab}\paraa{\d_af}\paraa{\d_bh},
\end{align}
where $\eps^{ab}$ is antisymmetric with $\eps^{12}=1$ and
$\d_a=\frac{\d}{\d u^a}$. With this bracket we define the tensors
\begin{align}
  &\P^{ij} = \{x^i,x^j\}\\
  &\SA^{ij} =\frac{1}{\rho}\eps^{ab}\paraa{\d_ax^i}\paraa{\nablab_bN_A}^j
  =\{x^i,n_A^j\}+\{x^i,x^k\}\Gammab^j_{kl}n_A^l,\label{eq:SA}
\end{align}
and one can also consider them as maps $TM\to TM$ by lowering the
second index with the ambient metric $\gb$, i.e.
\begin{align}
  &\P(X) = \P^{ik}\gb_{kj}X^j\d_i\\
  &\SA(X) = \SA^{ik}\gb_{kj}X^j\d_i.
\end{align}
With these definitions, one finds that
\begin{align}
    &\Tr\SA^2\equiv \paraa{\SA}^i_j\paraa{\SA}^j_i = -\frac{2}{\rho^2}\det(h_{A,ab})\\
    &\Tr\P^2\equiv \P^i_j\P^j_i = -2\frac{g}{\rho^2},\label{eq:trP2}  
\end{align}
where $g=\det\paraa{\gb(e_a,e_b)}$ is the determinant of the induced
metric on $\Sigma$, and $h_{A,ab}$ is the second fundamental form
corresponding to the normal vector $N_A$.

In the main part of this article we will make use of the
following result:

\begin{theorem}[\cite{ahh:psurface}]\label{thm:classicalK}
  Let $K$ denote the Gaussian curvature of $\Sigma$. Then
  \begin{align}
    &K = \frac{1}{g}\gb\paraa{\Rb(e_1,e_2)e_2,e_1}
    -\frac{\rho^2}{2g}\sum_{A=1}^p\Tr\SA^2,
  \end{align}
  where $\Rb$ is the curvature tensor of $M$.
\end{theorem}

\section{Matrix regularizations and discrete curvatures}

\noindent In the following, we shall assume that $\Sigma$ is a compact
closed orientable surface. Let us first define what is meant by a
``matrix regularization'', and then show some of its properties.

\begin{definition}
  Let $N_1,N_2,\ldots$ be a strictly increasing sequence of positive
  integers, let $\{\Ta\}$ for $\alpha=1,2,\ldots$ be linear maps from
  $C^\infty(\Sigma,\reals)$ to hermitian $\Na\times \Na$ matrices and let
  $\hbar(N)$ be a real-valued strictly positive decreasing function
  such that $\limNinf N\hbar(N)<\infty$. Furthermore, let $\omega$ be
  a symplectic form on $\Sigma$ and let $\{\cdot,\cdot\}$ denote the
  Poisson bracket induced by $\omega$. If $\{\Ta\}$ has the following
  properties for all $f,h\in C^\infty(\Sigma)$
  \begin{align}
    &\limainfty\norm{\Ta(f)}<\infty\label{eq:matrixNorm},\\
    &\limainfty\norm{\Ta(fh)-\Ta(f)\Ta(h)}=0,\label{eq:matrixProduct}\\
    &\limainfty\norm{\frac{1}{i\hbara}\com{\Ta(f),\Ta(h)}-\Ta\paraa{\pb{f,h}}}=0,\label{eq:matrixCommutator}\\
    &\limainfty 2\pi\hbara\Tr\Ta(f)=\int_\Sigma f \omega,\label{eq:matrixTrace}
  \end{align}
  where $||\cdot||$ denotes the operator norm and
  $\hbar_\alpha=\hbar(\Na)$, then we call the pair $(\Ta,\hbar)$ a
  \emph{matrix regularization of $(\Sigma,\omega)$}.
\end{definition}

\noindent Given local coordinates $u^1,u^2$ on $\Sigma$, we write $\omega
= \rho(u^1,u^ 2) du^1\wedge du^2$, and it is easy to see that the induced
Poisson bracket becomes
\begin{align*}
  \{f,h\} = \frac{1}{\rho}\eps^{ab}\paraa{\d_a f}\paraa{\d_b h}.
\end{align*}

\begin{definition}
  If $\fh_1,\fh_2,\ldots$ is a sequence of matrices such that
  $\fha$ has dimension $N_\alpha$ and if it holds that
  \begin{align}
    \limainfty\norm{\fha-\Ta(f)} = 0,
  \end{align}
  then we say that the sequence \emph{converges to the function $f$}.
\end{definition}

\begin{definition}
  A matrix regularization $(\Ta,\hbar)$ is called \emph{unital} if
  \begin{align}
    \limainfty\norm{\mid_{N_\alpha}-\Ta(1)}=0.
  \end{align}
\end{definition}

\begin{remark}
  Although unital matrix regularizations seem natural, and all our
  examples fall into this category, it is easy to construct examples
  of non-unital matrix regularizations. Namely, let $(\Ta,\hbar)$ be a
  matrix regularization and consider the map $\tilde{T}^\alpha$
  defined by
  \begin{align*}
    \tilde{T}^\alpha(f) = 
    \begin{pmatrix}
      &    &          &  & 0\\
      &    & \Ta(f)   &  & \vdots \\
      &    &          &  & \\
    0 &         & \cdots & & 0 
    \end{pmatrix}.
  \end{align*}
  Then $(\tilde{T}^\alpha,\hbar)$ is a matrix regularization which is not unital, since
  \begin{align*}
    \limainfty\norm{\tilde{T}^\alpha(1)-\mid_{N_\alpha+1}}\geq 1.
  \end{align*}
\end{remark}

\begin{proposition}
  Let $(\Ta,\hbar)$ be a unital matrix regularization. Then
  \begin{align}
    \limainfty 2\pi N_\alpha\hbara = \int_\Sigma\omega.
  \end{align}
\end{proposition}

\begin{proof}
  Let us use formula (\ref{eq:matrixTrace}) with $f=1$.
  \begin{align*}
    \int_\Sigma\omega &= \limainfty 2\pi\hbara\Tr\Ta(1)
    =\limainfty 2\pi\hbara\Tr\bracketb{\Ta(1)+\mid_{N_\alpha}-\mid_{N_\alpha}}\\
    &= \limainfty\parab{2\pi\hbara N_\alpha + 2\pi\hbara\Tr(\Ta(1)-\mid_{N_\alpha})}
    = \limainfty 2\pi\hbara N_\alpha
  \end{align*}
  since 
  \begin{align*}
    \limainfty \abs{2\pi\hbara\Tr(\Ta(1)-\mid_{N_\alpha})}
    \leq \limainfty 2\pi\hbara N_\alpha\norm{\Ta(1)-\mid_{N_\alpha}} = 0,
  \end{align*}
  due to the fact that the matrix regularization is unital.
\end{proof}

\begin{proposition}\label{prop:arbitrarySequences}
  Let $(\Ta,\hbar)$ be a matrix regularization of $(\Sigma,\omega)$
  and let $\{\fhua_k\}$ be a sequence of matrices converging to $f_k\in
  C^\infty(\Sigma)$ for $k=1,\ldots,n$. Then
  $\{a_1\fhua_1+\cdots+a_n\fhua_n\}$ converges to $a_1f_1+\cdots+a_nf_n$
  for any $a_1,\ldots,a_n\in\reals$ and
  \begin{align}
    &\limainfty\norm{\fhua_1\cdots\fhua_n} \leq \prod_{k=1}^n\limainfty\norm{\Ta(f_k)}\label{eq:multiMatrixNorm}\\
    &\limainfty\norm{\fhua_1\cdots\fhua_n-\Ta(f_1\cdots f_n)} = 0\label{eq:multiMatrixProduct}\\
    &\limainfty 2\pi\hbara\Tr\paraa{\fhua_1\cdots\fhua_n} = \int_{\Sigma}f_1\cdots f_n\omega\label{eq:multiMatrixTrace}.
  \end{align}
\end{proposition}

\begin{proof}
  The first statement about $\{a_1\fhua_1+\cdots a_n\fhua_n\}$ follows
  directly from the linearity of $\Ta$.  Let us prove
  (\ref{eq:multiMatrixNorm}) by induction on $n$. Thus, we assume that
  (\ref{eq:multiMatrixNorm}) holds and compute
  \begin{align*}
    \limainfty&\norm{\fhua_1\cdots\fhua_{n+1}} \leq 
    \limainfty\norm{\fhua_1\cdots\fhua_{n}}\norm{\fhua_{n+1}-\Ta(f_{n+1})+\Ta(f_{n+1})}\\
    &\leq\limainfty\Big(\norm{\fhua_{1}\cdots\fhua_n}\norm{\fhua_{n+1}-\Ta(f_{n+1})}
    +\norm{\fhua_1\cdots\fhua_n}\norm{\Ta(f_{n+1})}\Big)\\
    &=\limainfty\norm{\fhua_1\cdots\fhua_n}\norm{\Ta(f_{n+1})}
    \leq \prod_{k=1}^{n+1}\limainfty\norm{\Ta(f_k)}.
  \end{align*}
  To prove (\ref{eq:multiMatrixProduct}) we again proceed by induction
  and assume that (\ref{eq:multiMatrixProduct}) holds for any given
  $n$, and then compute
  \begin{align*}
    &\limainfty\norm{\fhua_1\cdots\fhua_{n+1}-\Ta(f_1\cdots f_{n+1})}\\
    &\leq\limainfty\Big(\norm{\fhua_1\cdots\fhua_n}\norm{\fhua_{n+1}-\Ta(f_{n+1})}
    +\norm{\fhua_1\cdots\fhua_{n}\Ta(f_{n+1})-\Ta(f_1\cdots f_{n+1})}\Big)\\
    &\leq\limainfty\Big(\norm{\fhua_1\cdots\fhua_n}\norm{\fhua_{n+1}-\Ta(f_{n+1})}
    +\Big|\Big|\fhua_1\cdots\fhua_{n}\Ta(f_{n+1})\\
      &\qquad\qquad-\Ta(f_1\cdots f_{n+1})-\Ta(f_1\cdots f_n)\Ta(f_{n+1})+\Ta(f_1\cdots f_n)\Ta(f_{n+1})\Big|\Big|\Big)\\
  &\leq\limainfty\Big(\norm{\fhua_1\cdots\fhua_n}\norm{\fhua_{n+1}-\Ta(f_{n+1})}
  +\norm{\Ta(f_{n+1})}\norm{\fhua_1\cdots\fhua_n-\Ta(f_1\cdots f_n)}\\
  &+\norm{\Ta(f_1\cdots f_n)\Ta(f_{n+1})-\Ta(f_1\cdots f_{n+1})}\Big)=0.
  \end{align*}
  Finally, we prove the trace formula:
  \begin{align*}
    \limainfty &2\pi\hbara\Tr\paraa{\fhua_1\cdots\fhua_n}\\
    &=\limainfty 2\pi\hbara
    \bracketb{\Tr\Ta(f_1\cdots f_n)+\Tr\paraa{\fhua_1\cdots\fhua_n-\Ta(f_1\cdots f_n)}}\\
    &=\int_{\Sigma}f_1\cdots f_n\omega,
  \end{align*}
  since 
  \begin{align*}
    \limainfty&\abs{\hbara\Tr\paraa{\fhua_1\cdots\fhua_n-\Ta(f_1\cdots f_n)}}\\
    &\leq \limainfty\hbara N_\alpha\norm{\fhua_1\cdots\fhua_n-\Ta(f_1\cdots f_n)}=0
  \end{align*}
  by formula (\ref{eq:multiMatrixProduct}).
\end{proof}

\noindent The above result allows one to easily construct sequences of
matrices converging to any sum of products of functions and Poisson
brackets. Namley, simply substitute for every factor in every term of
the sum, a sequence converging to that function, where Poisson
brackets of functions may be replaced by commutators of
matrices. Proposition \ref{prop:arbitrarySequences} then guarantees
that the matrix sequence obtained in this way converges to the sum of
the products of the corresponding functions.

\begin{proposition}
  Let $(\Ta,\hbar)$ be a matrix regularization and assume that
  $\{\fha\}$ converges to $f$. Then $\{\fha^\dagger\}$
  converges to $f$.
\end{proposition}

\begin{proof}
  Due to the fact that $||A||=||A^\dagger||$ one sees that
  \begin{align*}
    \limainfty\norm{\fha^\dagger-\Ta(f)}
   = \limainfty\norm{\paraa{\fha-\Ta(f)}^\dagger}
   = \limainfty\norm{\fha-\Ta(f)} = 0,
  \end{align*}
  since $\{\fha\}$ converges to $f$.
\end{proof}

\noindent If the matrix regularization is unital, one can relate the
matrix sequence converging to the function $1/f$, to the inverse of a
sequence converging to $f$.

\begin{proposition}
  Let $(\Ta,\hbar)$ be a unital matrix regularization and assume that
  $f$ is a nowhere vanishing function and that $\{\fha\}$
  converges to $f$. If $\fha^{-1}$ exists and
  $||\fha^{-1}||$ is uniformly bounded for all $\alpha$, then
  $\{\fha^{-1}\}$ converges to $1/f$.
\end{proposition}

\begin{proof}
  One calculates
  \begin{align*}
    \limainfty&\norm{\fha^{-1}-\Ta(1/f)}
    \leq \limainfty\norm{\fha^{-1}}\norm{\mid_{N_\alpha}-\fha\Ta(1/f)}\\
    &=\limainfty\norm{\fha^{-1}}\norm{\mid_{N_\alpha}-\fha\Ta(1/f)+\Ta(1)-\Ta(1)}\\
    &\leq\limainfty\norm{\fha^{-1}}\parab{\norm{\mid_{N_\alpha}-\Ta(1)}
      +\norm{\fha\Ta(1/f)-\Ta(1)}}\\
    &=0,
  \end{align*}
  since the matrix regularization is unital and $||\fha^{-1}||$
  is assumed to be uniformly bounded.
\end{proof}

\noindent Recall that $\Sigma$ is embedded in a $m=2+p$ dimensional manifold $M$
via the embedding coordinates $x^1(u^1,u^2),\ldots,x^m(u^1,u^2)$, and
that $p$ orthonormal normal vectors are given with components $n_A^i$.  By
$\{X^i_\alpha\}$ and $\{N_{A,\alpha}^i\}$ we will denote arbitrary
sequences converging to $x^i$ and $n_A^i$ respectively. Moreover,
given the metric $\gb_{ij}$ and the Christoffel symbols
$\Gammab^{i}_{jk}$ of $M$, we let $\{\Gh_{ij,\alpha}\}$ and
$\{\Gammah^{i}_{jk,\alpha}\}$ denote sequences converging to
$\gb_{ij}$ and $\Gamma^{i}_{jk}$ respectively. To avoid excess of
notation, we suppress the index $\alpha$ whenever all matrices are
considered at a fixed (but arbitrary) $\alpha$.

In analogy with (\ref{eq:SA}) we define
\begin{align}
  (\ShA)^j_k = \frac{1}{i\hbar}[X^j,N_A^{k'}]\Gh_{k'k}+\frac{1}{i\hbar}[X^j,X^l]\Gammah^{k'}_{lm}N_A^m\Gh_{k'k},
\end{align}
and 
\begin{align}
  \trh\ShA^2 = \paraa{(\ShA)^j_k}^\dagger(\ShA)^k_j.
\end{align}

\noindent Let $g_{ab}$ be the induced metric on $\Sigma$, and $g$ its determinant. We set
\begin{align}
  \gamma = \frac{\sqrt{g}}{\rho},
\end{align}
and denote by $\{\gammah_\alpha\}$ an arbitrary sequence of invertible matrices converging to $\gamma$.
By defining
\begin{align*}
  \Ph^j_k = \frac{1}{i\hbar}[X^j,X^l]\Gh_{lk},
\end{align*}
it follows from (\ref{eq:trP2}) that
\begin{align}
  -\frac{1}{2}(\Ph^i_k)^\dagger\Ph^k_i = \frac{1}{2\hbar^2}\Gh_{jk}^\dagger[X^i,X^j][X^k,X^l]\Gh_{li}
\end{align}
converges to $\gamma^2$.

If the embedding space is $\reals^m$, the above formulas reduce to
\begin{align}
  &\trh\ShA^2 = -\frac{1}{\hbar^2}\sum_{i,j=1}^m[X^i,N_A^j][X^j,N_A^i],\\
  -\frac{1}{2}&(\Ph^i_k)^\dagger\Ph^k_i = -\frac{1}{\hbar^2}\sum_{i<j}^m[X^i,X^j]^2,
\end{align}
and in $\reals^3$ one obtains 
\begin{align}\label{eq:trSreals3}
  \trh\Sh^2 = \frac{1}{4\hbar^4}\sum^3
  \eps_{jkl}\eps_{ik'l'}(\gammah^\dagger)^{-1}\bracketa{X^i,[X^k,X^l]}\bracketa{X^j,[X^{k'},X^{l'}]}\gammah^{-1},
\end{align}
since 
\begin{align}
  n^i = \frac{1}{2\gamma}\eps^i_{jk}\{x^j,x^k\},
\end{align}
defines a unit normal vector to the surface (cp. \cite{ahh:psurface},
where (\ref{eq:trSreals3}) is also given for arbitrary codimension).

We are now ready to define and present formulas for the discrete
curvature in a matrix regularization of $\Sigma$.

\begin{definition}
  Let $(\Ta,\hbar)$ be a matrix regularization of $(\Sigma,\omega)$
  and let $K$ be the Gaussian curvature of $\Sigma$. A \emph{Discrete
    Curvature of $\Sigma$} is a matrix sequence
  $\{\Kh_1,\Kh_2,\Kh_3,\ldots\}$ converging to $K$,
  and a \emph{Discrete Euler Characteristic of $\Sigma$} is a sequence
  $\{\chih_1,\chih_2,\chih_3,\ldots\}$ such that
  $\displaystyle\limainfty\chih_\alpha=\chi$.
\end{definition}

\noindent From the classical Gauss-Bonnet theorem, it is immediate to derive a
discrete analogue for matrix regularizations.

\begin{theorem}\label{thm:discreteEuler}
  Let $(\Ta,\hbar)$ be a matrix regularization of $(\Sigma,\omega)$, and let
  $\{\Kh_1,\Kh_2,\ldots\}$ be a discrete curvature of $\Sigma$. Then the sequence
  $\chih_1,\chih_2,\ldots$ defined by
  \begin{align}
    \chih_{\alpha} = \hbara\Tr\bracketb{\gammah_\alpha\Kh_{\alpha}},
  \end{align}
  is a discrete Euler characteristic of $\Sigma$.
\end{theorem}

\begin{proof}
  To prove the statement, we compute $\limainfty \chih_\alpha$ and show that it is equal to $\chi(\Sigma)$. Thus
  \begin{align*}
    \limainfty\chih_\alpha &= \limainfty\frac{1}{2\pi}2\pi\hbara\Tr\bracketb{\gammah_\alpha\Kh_{\alpha}},
  \end{align*}
  and by using Proposition \ref{prop:arbitrarySequences} we can write
  \begin{align*}
    \limainfty\chih_\alpha &= \frac{1}{2\pi}\int_{\Sigma}K\frac{\sqrt{g}}{\rho}\omega
    =\frac{1}{2\pi}\int_{\Sigma}K\frac{\sqrt{g}}{\rho}\rho dudv =
    \frac{1}{2\pi}\int_{\Sigma}K\sqrt{g}dudv = \chi(\Sigma),
  \end{align*}
  where the last equality is the classical Gauss-Bonnet theorem.
\end{proof}

\begin{theorem}\label{thm:discreteCurvature}
  Let $(\Ta,\hbar)$ be a unital matrix regularization of $(\Sigma,\omega)$
  and let $\{\bar{K}_{12}^\alpha\}$ be a matrix sequence converging to
  $\gb\paraa{\Rb(e_1,e_2)e_2,e_1}/g$ (the sectional curvature of
  $\TSigma$ in $M$). Then the sequence $\Kh$ of matrices defined by
  \begin{align}
    \Kh = \bar{K}_{12}-\frac{1}{2}\sum_{A=1}^p(\gammah^\dagger)^{-1}\paraa{\trh\ShA^2}\gammah^{-1}
  \end{align}
  is a discrete curvature of $\Sigma$. Thus, a discrete Euler
  characteristic is given by
  \begin{align}
    \chih = \hbar\Tr\paraa{\gammah\bar{K}_{12}}-\frac{\hbar}{2}\sum_{A=1}^p\Tr\bracketb{\gammah^{-1}\trh\ShA^2}.
  \end{align}
\end{theorem}

\begin{proof}
  By using the way of constructing matrix sequences given through
  Proposition \ref{prop:arbitrarySequences}, the result follows
  immediately from Theorem \ref{thm:classicalK}.
\end{proof}

\noindent Note that if $\rho=\sqrt{g}$, then $\gamma=1$ which implies
that one can choose $\gammah_\alpha=\mid_{N_\alpha}$ when the matrix regularization is
unital.

\section{Two simple examples}

\subsection{The round fuzzy sphere}

\noindent For the sphere embedded in $\reals^3$ as 
\begin{align}
  \xv = (x^1,x^2,x^3) = (\cos\vphi\sin\theta,\sin\vphi\sin\theta,\cos\theta)
\end{align}
with the induced metric
\begin{align}
  (g_{ab}) = 
  \begin{pmatrix}
    1 & 0 \\ 0 & \sin^2\theta
  \end{pmatrix},
\end{align}
it is well known that one can construct a matrix regularization from
representations of $su(2)$. Namely, let $S_1,S_2,S_3$ be hermitian
$N\times N$ matrices such that $[S^j,S^k] = i{\epsilon^{jk}}_lS^l$,
$(S^1)^2+(S^2)^2+(S^3)^2=(N^2-1)/4$, and define
\begin{align}
  X^i = \frac{2}{\sqrt{N^2-1}}S^i.
\end{align}
Then there exists a map $\TN$ (which can be defined through expansion
in spherical harmonics) such that $\TN(x^i)=X^i$ and
$(\TN,\hbar=2/\sqrt{N^2-1})$ is a unital matrix regularization of
$(S^2,\sqrt{g}d\theta\wedge d\vphi)$ \cite{h:phdthesis}.  A unit normal of the sphere in
$\reals^3$ is given by $N\in T\reals^3$ with $N=x^i\d_i$, which gives
$N^i=X^i$, and one can compute the discrete curvature as
\begin{align}
  \Kh_N = -\frac{1}{\hbar^2}\sum_{i<j=1}^m\Tr[X^i,X^j]^2 = \mid_N 
\end{align}
which gives the discrete Euler characteristic
\begin{align}
  \chih_N &= \hbar\Tr\Kh_N = \hbar N = \frac{2N}{\sqrt{N^2-1}},
\end{align}
converging to $2$ as $N\to\infty$.

\subsection{The fuzzy Clifford torus}

\noindent The Clifford torus in $S^3$ can be regarded as embedded in $\reals^4$ through
\begin{align*}
  \xv = (x^1,x^2,x^3,x^4) = \frac{1}{\sqrt{2}}(\cos\vphi_1,\sin\vphi_1,\cos\vphi_2,\sin\vphi_2),
\end{align*}
with the induced metric
\begin{align*}
  (g_{ab}) = \frac{1}{2}
  \begin{pmatrix}
    1 & 0 \\ 0 & 1
  \end{pmatrix},
\end{align*}
and two orthonormal vectors, normal to the tangent plane of the
surface in $T\reals^4$, can be written as
\begin{align*}
  N_\pm = x^1\d_1 + x^2\d_2 \pm x^3\d_3  \pm x^4\d_4.
\end{align*}
To construct a matrix regularization for the Clifford torus, one
considers the $N\times N$ matrices $g$ and $h$ with non-zero elements
\begin{align*}
  &g_{kk} = \omega^{k-1}\quad\text{ for $k=1,\ldots,N$}\\
  &h_{k,k+1} = 1\quad\text{ for $k=1,\ldots,N-1$}\\
  &h_{N,1} = 1,
\end{align*}
where $\omega=\exp(i2\theta)$ and $\theta=\pi/N$. These matrices satisfy
the relation $hg=\omega gh$.  The map $\TN$ is then defined on the
Fourier modes
\begin{align*}
  Y_{\mv}=e^{i\mv\cdot\vphiv}=e^{im_1\vphi_1+im_2\vphi_2}
\end{align*}
as  
\begin{align*}
  \TN(Y_{\mv}) = \omega^{\frac{1}{2}m_1m_2}g^{m_1}h^{m_2},
\end{align*}
and the pair $(\TN,\hbar=\sin\theta)$ is a unital matrix
regularization of the Clifford torus with respect to
$\sqrt{g}d\vphi_1\wedge d\vphi_2$
\cite{ffz:trigonometric,h:diffeomorphism}. Thus, using this map one
finds that
\begin{align*}
  &X^1 = T(x^1) = \frac{1}{\sqrt{2}}T(\cos\vphi_1) = \frac{1}{2\sqrt{2}}(\gd+g)\\
  &X^2 = T(x^2) = \frac{1}{\sqrt{2}}T(\sin\vphi_1) = \frac{i}{2\sqrt{2}}(\gd-g)\\
  &X^3 = T(x^3) = \frac{1}{\sqrt{2}}T(\cos\vphi_2) = \frac{1}{2\sqrt{2}}(\hd+h)\\
  &X^4 = T(x^4) = \frac{1}{\sqrt{2}}T(\sin\vphi_2) = \frac{i}{2\sqrt{2}}(\hd-h)
\end{align*}
which implies that $N_\pm^1=X^1$, $N_\pm^2=X^2$, $N_\pm^3=\pm X^3$ and
$N_\pm^4=\pm X^4$. By a straightforward computation one obtains
\begin{align*}
  -\frac{1}{\hbar^2}\sum_{i,j=1}^4[X^i,X^j]^2 = 2\mid
\end{align*}
and therefore
\begin{align*}
  \frac{1}{2\hbar^2}\sum_{i,j=1}^4[X^i,N^j_+][X^j,N^i_+]=-\frac{1}{2\hbar^2}\sum_{i,j=1}^4[X^i,X^j]^2 = \mid,
\end{align*}
and since $[X^1,X^2]=[X^3,X^4]=0$ it follows that
\begin{align*}
  \frac{1}{2\hbar^2}\sum_{i,j=1}^4[X^i,N^j_-][X^j,N^i_-]
  =\frac{1}{2\hbar^2}\sum_{i,j=1}^4[X^i,X^j]^2 = -\mid.
\end{align*}
This implies that the discrete curvature vanishes, i.e.
\begin{align*}
  \Kh_N = \frac{1}{2\hbar^2}\sum_{i,j=1}^4[X^i,N^j_+][X^j,N^i_+]
  +\frac{1}{2\hbar^2}\sum_{i,j=1}^4[X^i,N^j_-][X^j,N^i_-] = \mid-\mid = 0,
\end{align*}
which immediately gives $\chih_N=0$.

\section{Axially symmetric surfaces in $\reals^3$}\label{sec:axiallysymmetric}

\noindent Recall the classical description of general axially symmetric surfaces:
\begin{align}\label{axiallyuvparam}
  \xv &= \paraa{f(u)\cos v, f(u)\sin v, h(u)}\\
  \nv &= \frac{\pm 1}{\sqrt{h'(u)^2+f'(u)^2}}
  \paraa{h'(u)\cos v,h'(u)\sin v,-f'(u)}\notag,
\end{align}
which implies
\begin{align*}
  \paraa{g_{ab}}=
  \begin{pmatrix}
    f'^2+h'^2 & 0 \\
    0 & f^2
  \end{pmatrix}\qquad
  \paraa{h_{ab}} =\frac{\pm 1}{\sqrt{h'^2+f'^2}}
  \begin{pmatrix}
    h'f''-h''f' & 0\\
    0 & -fh'
  \end{pmatrix},
\end{align*}
where $h_{ab}$ are the components of the second fundamental form. The
Euler characteristic can be computed as
\begin{align}
  \chi = \frac{1}{2\pi}\int K\sqrt{g} = 
  -\int_{u_-}^{u_+}\frac{h'\paraa{h'f''-h''f'}}{\paraa{f'^2+h'^2}^{3/2}}du
  =-\frac{f'}{\sqrt{f'^2+h'^2}}\Bigg|_{u_-}^{u_+},
\end{align}
which is equal to zero for tori (due to periodicity) and equal to $+2$ for spherical surfaces (due to $f'(u_{\pm})=\mp\infty$).

While a general procedure for constructing matrix analogues of
surfaces embedded in $\reals^3$ was obtained in
\cite{abhhs:noncommutative,abhhs:fuzzy} (cp. also \cite{a:repcalg}),
let us restrict now to $h(u)=u=z$, hence describe the axially
symmetric surface $\Sigma$ as a level set, $C=0$, of
\begin{align}
  C(\xv) = \frac{1}{2}\paraa{x^2+y^2-f^2(z)},
\end{align}
to carry out the construction in detail, and make the resulting
formulas explicit. Defining 
\begin{align}
  \{F(\xv),G(\xv)\}_{\reals^3} = \nabla C\cdot\paraa{\nabla F\times\nabla G},
\end{align}
one has 
\begin{align}
  \{x,y\}=-\ff'(z),\quad\{y,z\}=x,\quad\{z,x\} = y,
\end{align}
respectively
\begin{align}\label{eq:XYZCommutators}
  [X,Y] = i\hbar \ff'(Z),\quad [Y,Z]=i\hbar X,\quad [Z,X]=i\hbar Y
\end{align}
for the ``quantized'' (``non-commutative'') surface. In terms of the parametrization given in 
(\ref{axiallyuvparam}), the above Poisson bracket is equivalent to
\begin{align}
  \{F(u,v),G(u,v)\} = \eps^{ab}\paraa{\d_aF}\paraa{\d_b{G}}
\end{align}
where $\d_1=\d_v$ and $\d_2=\d_u$. By finding matrices of increasing
dimension satisfying (\ref{eq:XYZCommutators}), one can construct a
map $\Ta$ having the properties (\ref{eq:matrixProduct}) and
(\ref{eq:matrixCommutator}) of a matrix regularization restricted to
polynomial functions in $x,y,z$ (cp. \cite{a:phdthesis}).

For the round 2-sphere, $f(z)=1-z^2$, (\ref{eq:XYZCommutators}) gives
the Lie algebra $su(2)$, and its celebrated irreducible
representations satisfy
\begin{align}\label{eq:su2sumsquare}
  X^2+Y^2+Z^2 = \mid\quad\text{if}\quad \hbar=\frac{2}{\sqrt{N^2-1}}.
\end{align}
When $f$ is arbitrary, one can still find finite dimensional
representations of (\ref{eq:XYZCommutators}) as follows: rewrite
(\ref{eq:XYZCommutators}) as
\begin{align}
  &[Z,W] = \hbar W\label{eq:ZWCommutator}\\
  &[W,\Wd] = -2\hbar\ff'(Z)
\end{align}
implying that $z_i-z_j=\hbar$ whenever $W_{ij}\neq 0$ and $Z$
diagonal. Assuming $W=X+iY$ with non-zero matrix elements
$W_{k,k+1}=w_k$ for $k=1,\ldots,N-1$, one thus obtains (with
$w_0=w_N=0$)
\begin{align*}
  &Z_{kk} = \frac{\hbar}{2}\paraa{N+1-2k}\\
  &w_k^2-w_{k-1}^2=-2\hbar\ff'\paraa{\hbar(N+1-2k)/2}\equiv Q_k,
\end{align*}
which implies that
\begin{align*}
  w_ k^2 = \sum_{l=1}^kQ_l
\end{align*}
and the only non-trivial problem is to find the analogue of
(\ref{eq:su2sumsquare}). To this end, define 
\begin{align}\label{eq:fhdef}
  \fh^2 = X^2+Y^2 = \frac{1}{2}\paraa{W\Wd+\Wd W},
\end{align}
with $W$ given as above. As $Z$ has pairwise different eigenvalues,
the diagonal matrix given in (\ref{eq:fhdef}) can be thought of as a
function of $Z$; hence as $\fh^2(Z)$. It then trivially holds that
\begin{align}
  \Ch = X^2+Y^2-\fh^2(Z)=0,
\end{align}
for the representation defined above. The quantization of $\hbar$
comes through the requirement that $\fh^2$ should correspond to
$f^2$. While for the \emph{round} 2-sphere $\fh^2$ equals $f^2$,
provided $\hbar$ is chosen as in (\ref{eq:su2sumsquare}), it is easy
to see that in general they can not coincide, as
\begin{align*}
  [X^2+Y^2-&f(Z)^2,W] = [(W\Wd+\Wd W)/2-f(Z)^2,W]\\
  &=\frac{1}{2}W[\Wd,W]+\frac{1}{2}[\Wd,W]W-f(Z)[f(Z),W]-[f(Z),W]f(Z)\\
  &=\cdots=f(Z)\paraa{\hbar f'(Z)W-[f(Z),W]}+\paraa{\hbar f'(Z)W-[f(Z),W]}f(Z)
\end{align*}
with off-diagonal elements
\begin{align*}
  \paraa{f(z_k)+f(z_{k-1})}\paraa{\hbar f'(z_k)-(f(z_k)-f(z_{k-1}))}
\end{align*}
that are in general non-zero (hence $X^2+Y^2+f^2(Z)$ is usually not
even a Casimir, except in leading order).

How it \emph{does} work is perhaps best illustrated by a non-trivial example, $f(z)=1-z^4$:
\begin{align}
    w_k^2 =\frac{\hbar^4}{2}&\parab{(N+1)^3k-3(N+1)^2k(k+1)+\label{eq:wk}\\
      &2(N+1)k(k+1)(2k+1)-2k^2(k+1)^2}\notag\\
  \fh_k^2 = \frac{1}{2}(w_k^2&+w^2_{k-1}) = 
  \frac{\hbar^4}{4}\parab{(N+1)^3(2k-1)-6(N+1)^2k^2\notag\\ &\qquad\qquad+4(N+1)k(2k^2+1)-4k^2(k^2+1)}\notag
\end{align}
(note that $w^2_0=w_N^2=0$ is explicit in (\ref{eq:wk})) so that
\begin{align}
  \paraa{X^2+Y^2+Z^4}_{kk} = \hbar^4\bracketc{\frac{(N+1)^4}{16}-\frac{(N+1)^3}{4}+k(N+1)-k^2}.
\end{align}
Expressing the last two terms via $Z^2$ (note that the cancellation of
$k^3$ and $k^4$ terms shows the absence of $Z^3$ and higher
corrections) one finds
\begin{align*}
  X^2+Y^2+Z^4+\hbar^2Z^2 &= \hbar^4\frac{(N+1)^2}{16}\parab{(N+1)^2-4(N+1)+4}\mid\\
  &=\hbar^4\frac{(N^2-1)^2}{16}\mid,
\end{align*}
which equals $\mid$ if $\hbar$ is chosen as $2/\sqrt{N^2-1}$. Note
that this is the \emph{same} expression for $\hbar$ then for the round
sphere, $f^2=1-z^2$ (cp. (\ref{eq:su2sumsquare})).

A more elegant way to derive the quantum Casimir (cp. also \cite{r:repnonlinear,gps:beyondfuzzy})
\begin{align}
  Q = X^2+Y^2+Z^4+\hbar^2Z^2
\end{align}
is to calculate
\begin{align*}
  [X^2+Y^2+Z^4,W] &= [(W\Wd+\Wd W)/2+Z^4,W]\\
  &= \cdots = \hbar^2[W,Z^2],
\end{align*}
which determines the terms proportional to $\hbar$ in the Casimir. 

Due to the general formula 
\begin{align}
  \Kh = -\frac{1}{8\hbar^4}\eps_{jkl}\eps_{ipq}(\gammah^\dagger)^{-2}\coma{X^i,[X^k,X^l]}\coma{X^j,[X^p,X^q]}\gammah^{-2}
\end{align}
one obtains, for the axially symmetric surfaces discussed above,
\begin{align}
  \Kh = \gammah^{-2}\parac{(\ff')^2(Z)+\frac{1}{2\hbar}[W,\ff'(Z)]\Wd+\frac{1}{2\hbar}\Wd[W,\ff'(Z)]}\gammah^{-2}
\end{align}
with
\begin{align}
  \gammah^2 = \frac{1}{2}\paraa{W\Wd+\Wd W}+(\ff')^2(Z)
  =f(Z)^2\paraa{f'(Z)^2+\mid} + O(\hbar),
\end{align}
giving
\begin{align}
  &\Kh  = -\paraa{f'(Z)^2+\mid}^{-2}f(Z)^{-1}f''(Z) + O(\hbar)
\end{align}
and for $f(z)^2=1-z^4$ one has 
\begin{align}
  &\Kh = \paraa{4Z^6+\mid-Z^4}^{-2}\paraa{6Z^2-2Z^6}+O(\hbar)\\
  &\gammah^2 = \mid-Z^4+4Z^6+O(\hbar).
\end{align}
Note that (cp. (\ref{eq:ZWCommutator}))
$z_j-z_{j-1}=\hbar$ for arbitrary $f$, and that (due to the axial
symmetry) $\Kh$ and $\gammah^2$ are \emph{diagonal} matrices, so that
\begin{align*}
  \chih = \hbar\Tr\paraa{\sqrt{\gammah^2}\Kh},
\end{align*}
in this case simply being a Riemann sum approximation of $\int
K\sqrt{g}$, indeed converges to 2, the Euler characteristic of
spherical surfaces.

\section*{Acknowledgement}

\noindent J.H. would like to thank M. Bordemann for discussions, the
Albert Einstein Institute for hospitality, and D. O'Connor for the
November 2009 DIAS Workshop on Noncommutativity and Matrix Models,
where part of the present work was presented\footnote{J.H. ``The
  topology of non-commutative surfaces'', DIAS November
  2009.}. Furthermore, J.A. would like to thank H. Shimada for
discussions on matrix regularizations.

\bibliographystyle{alpha}
\bibliography{discretegb}

\end{document}